\documentclass[conference]{sig-alternate-10pt}
\usepackage{color}
\usepackage{algorithm,algorithmicx,algcompatible}
\usepackage{verbatim, subfigure}
\usepackage{array, calc, multicol}
\usepackage{xspace}
\usepackage{enumerate}
\usepackage{graphics, epstopdf, graphicx, epsfig, psfrag, epsf, subfigure}
\usepackage{amssymb, amsfonts, amsmath, times}
\usepackage{multicol, multirow, balance, verbatim,caption}
\usepackage[mediumspace,mediumqspace,Grey,squaren]{SIunits}

\usepackage{authblk}

\usepackage{amsmath, amsthm, amssymb}

\usepackage{relsize}

\usepackage{url}
\usepackage{breakurl}

\makeatletter
\def\url@leostyle{%
  \@ifundefined{selectfont}{\def\UrlFont{\sf}}{\def\UrlFont{\small\ttfamily}}}
\makeatother
\usepackage{textcomp}
\newcommand{\customtilde}{\raisebox{0.3ex}{\texttildelow}}

\newcommand{\eg}{{\it e.g.}\xspace}
\newcommand{\ie}{{\it i.e.}\xspace}
\newcommand{\etc}{{\it etc.}\xspace}

\newcommand\textvtt[1]{{\normalfont\fontfamily{cmvtt}\selectfont #1}}

\newcommand{\meddle}{\textvtt{Meddle}\xspace}

\newcommand{\haystack}{\textvtt{Hay\-stack}\xspace}

\newcommand{\recon}{\textvtt{Recon}\xspace}

\newcommand{\preDefLeak}{{\em predefined}\xspace}
\newcommand{\unknownLeak}{{\em unknown}\xspace}
\newcommand{\antShield}{\textvtt{Ant\-Shield}\xspace}
\newcommand{\antLib}{\textvtt{Ant\-Monitor Library}\xspace}

\newcommand{\anteater}{\textvtt{Ant\-Monitor}\xspace}
\newcommand{\antshield}{\textvtt{Ant\-Shield}\xspace}
\newcommand{\zeptolab}{\textvtt{com.ze\-p\-to\-lab.ctr.ads}\xspace}
\newcommand{\workaround}{\textvtt{work-a\-round}\xspace}
\newcommand{\taintdroid}{\textvtt{Taint\-Droid}\xspace}
\newcommand{\appfence}{\textvtt{App\-Fence}\xspace}

\newcommand{\stringMatch}{\textvtt{String Matching}\xspace}
\newcommand{\multiLabel}{\textvtt{Multi-Label}\xspace}

\begin{document}

\title{AntShield: On-Device Detection \\ of Personal Information Exposure}
\author[1]{Anastasia Shuba}
\author[1]{Evita Bakopoulou}
\author[1]{Milad Asgari Mehrabadi}
\author[1]{Hieu Le}
\author[2]{David Choffnes}
\author[1]{Athina Markopoulou}
\affil[1]{University of California, Irvine}
\affil[2]{Northeastern University}

\maketitle

\begin{abstract}
Mobile devices have access to personal, potentially sensitive data, and there is a growing number of applications that transmit this personally identifiable information (PII) over the network.  In this paper, we present the \antshield system that performs {\em on-device} packet-level monitoring  and detects the transmission of such sensitive information accurately and in real-time. A key insight is to distinguish PII that is \preDefLeak 
and is easily available on the device from PII that is \unknownLeak a priori but can be automatically detected by classifiers. Our system not only combines, for the first time, the advantages of {\em on-device} monitoring  
 with the power of  {\em learning}  unknown PII,  but also outperforms either of the two approaches alone. We demonstrate the real-time performance of our prototype as well as the classification performance using a dataset that we collect and analyze from scratch (including new findings in terms of leaks and patterns). \antshield is a first step towards enabling distributed learning of private information exposure.
\end{abstract}

\section{Introduction}

Mobile devices  have access to a wealth of personal, potentially sensitive information and there is a growing number of applications that access, process and transmit some of this information over the network. Sometimes this is justified (required for the intended operation of the applications, \eg location is needed by GoogleMaps) and controllable (\eg by the user  through permissions),  but for the most part,  users are not in control of their data today. %
 Applications and third party libraries routinely transmit user data to remote servers, including adservers and trackers, and users are typically unaware of how their personal data is shared and for what purpose. %

 Prior work on improving data transparency and identifying potential privacy leaks includes static and dynamic analysis and network-centric approaches. %
  In this paper, we take the latter approach: personal information  leaks happen, by definition, over network traffic, therefore a natural and comprehensive vantage point to identify and control leaks is at the network layer. Traffic can be monitored in the middle of the network (as in \meddle  \cite{meddle} and \recon \cite{recon15})  and/or on the device  itself (as in  \anteater \cite{antmonitor-arxiv}\cite{antmonitor-poster-mobicom15} and \textvtt{Lumen}  (a.k.a.  \haystack) \cite{razaghpanah2016haystack}).   A key challenge for network-based monitoring is how to analyze traffic both efficiently and securely.  The current state-of-the-art consists of the following complementary approaches. On one hand, \anteater \cite{antmonitor-arxiv} and  \textvtt{Lumen}  \cite{razaghpanah2016haystack} detect leaks on the device, but require a blacklist of strings (potential PII leaks) known a priori to search for. Therefore, they are unable to detect leakage of information that changes dynamically or is not part of the list.  On the other hand, \recon \cite{recon15} recently addressed this limitation, by training classifiers in a fully centralized way. However, the implementation relied on   %
 a trusted, remote proxy to route and analyze traffic, which potentially impacts scalability and security. %
 
We adopt the on-device network monitoring paradigm, which presents both opportunities and challenges. On the upside, it  obviates the need for a trusted infrastructure and gives full control to the user, which we believe is the right approach in privacy. Devices also have access to important contextual information, such as certain personal information available on the phone, and which apps are responsible for transmitting packets.
On the downside, mobile devices 
 have limited resources to conduct traffic analysis, including deep packet inspection (DPI),  %
 and training and applying machine learning classifiers for inferring leaks of PII. %
  It is currently an open question as to how to train machine learning classifiers to retain high accuracy in a truly distributed manner. 
  
In this paper, we take the first step towards enabling  distributed learning of personal information leaks from network traffic.  We present \antshield\xspace - a system that performs efficient on-device analysis,  provides accurate and comprehensive data privacy protection, and gives users  transparency and control over their personal information in real-time. 
  A key insight  is the distinction between PII that is \preDefLeak by the user or is readily available on the device, from PII  that is a priori \unknownLeak and should be  inferred by classifiers.  We  propose a hybrid  \stringMatch-classification approach:  %
  (i) we build on  the \antLib \cite{antmonitor-arxiv} for intercepting packets on the device and looking for \preDefLeak strings in real-time and (ii) we build classifiers 
  for the remaining \unknownLeak PII. 

The contributions of this paper are the following:
\begin{itemize}

 	\item 
the {\em \antshield System.} We present the first system to detect PII  exposure (using a {\em hybrid}  DPI and classification approach), 100\% on the device (from user space and without routing traffic through a remote VPN server), and  in real-time (in \customtilde1 ms). This is enabled by our system design and multiple optimizations. %
 	
 	\item 
	{\em Classification Methodology.} %
	Our
	{\em multi-label} classification methodology (Binary Relevance with Decision Trees) achieves significantly higher accuracy (8-25\% improvement) and lower variance (a factor of 2-5) compared to state-of-the-art.  %
	We also design and advocate for {\em per-app}, instead of {\em per-domain}, classifiers: they achieve similar classification accuracy, but allow faster and more scalable operation while covering more traffic.      
 	\item 
	{\em Dataset and Analysis.} In order to demonstrate the effectiveness of our approach, we collect a new (larger and richer than previously available) dataset of privacy leaks on mobile devices, which we will make available to the community. As a side contribution, we analyzed the dataset, identified previously unseen leaks (including leaks over plain TCP and UDP, leaks while the app is in the background, and malicious scanning for rooted devices) and behavioral patterns (\eg communities of domains and mobile apps involved in exposing private information). %
 	
\end{itemize} 

The structure of the rest of the  paper is as follows. Section \ref{sec:related} briefly reviews related work.
Section \ref{sec:system} describes our system's  rationale, design and implementation. %
 Section \ref{sec:eval} evaluates \antshield's classification accuracy and run-time performance; it also presents our collected dataset and findings therein.  Section \ref{sec:conclusion} concludes  the paper. 
\section{Related Work} \label{sec:related}

Different communities are working on improving data transparency and exposing or preventing potential privacy leaks. {\em Permissions} are useful but not sufficient: (i) users typically accept to install apps by default; (ii) permissions do not protect against inter-app  communication and  poorly documented system calls; and (iii) they do not capture run-time behavior. Using a {\em custom OS}  or a rooted phone one can get access to fine-grained information on the device, including network traffic.  \textvtt{Phonelab} \cite{phonelab} and others \cite{vallina2013rilanalyzer, wei2012profiledroid} use packet capturing APIs such as \textvtt{tcpdump} or \textvtt{iptables-log}. These are powerful  but  inherently limited to small scale-deployment as the overwhelming majority of  users do not have rooted phones,  and wireless providers and phone manufacturers strongly discourage rooting. The same limitation applies to approaches that use a custom OS to dynamically intercept leaks (\eg\xspace \textvtt{\taintdroid} \cite{enck2014taintdroid}) or permission requests to certain resources (\eg\xspace \textvtt{\appfence} \cite{hornyack2011thesedroids}).
Static analysis tools such as
\textvtt{AndroidLeaks} \cite{gibler2012androidleaks} and \textvtt{PiOS} \cite{egele2011pios} are limited by having to constantly download and analyze all available apps, which is not scalable. Moreover, static analysis suffers from inherent imprecisions, is not representative of what can happen at run-time, and cannot deal with native or dynamically loaded code.

Within the network measurements community, a number of prior works \cite{meddle, recon15, antmonitor-arxiv, antmonitor-poster-mobicom15, razaghpanah2016haystack}  %
have looked for personal information  leaks in network traffic. %
 This includes monitoring in the middle of the network (as in \meddle  \cite{meddle} and \recon \cite{recon15})  or on the device  itself (as in  \anteater (\cite{antmonitor-arxiv, antmonitor-poster-mobicom15}) and \textvtt{Lumen} (a.k.a. \haystack) \cite{razaghpanah2016haystack}).  \anteater and  \textvtt{Lumen} detect leaks on the device, but require a blacklist of strings (potential PII leaks) known a priori to search for; therefore, they are unable to detect leakage of information that changes dynamically or is not part of the list. 
 
To remedy this limitation, \recon recently applied machine learning techniques  to predict whether or not a given packet contains PII  \cite{recon15}. They broke packets into words based on delimiters (\eg `?', `=', `:') and then used these words as features in classification. Various methods were used to ensure that the PII themselves and strings that occur too often or too infrequently are not part of the feature list, see \cite{recon15} for details.
 To decide whether or not a packet contains PII (a binary classification problem), \recon used the Java Weka library's \cite{hall2009weka} C4.5 Decision Tree (DT), and then heuristics for extracting the type of leak. To improve classification accuracy \recon built specialized classifiers for each destination domain that received enough data to train such a classifier. For the rest of the domains, a {\em general} classifier was built. For the heuristic step, \recon maintained a list of probabilities that a particular key-word corresponds to a PII value. For each PII type, the probability was calculated by taking the number of times the key was present in a packet with the given PII, and dividing it by the number of times the key appeared in all packets. During PII extraction, \recon looked for keys with probability higher than an empirically computed threshold. Their code and dataset are available at \cite{recondata}.

 \recon is the closest to our work, thus we use it as our baseline for comparison throughout the paper.  The  key difference lies in the centralized vs distributed approach. \recon collected its datasets in the middle of the network, trained and applied the classifiers in a centralized way. \antShield operates on the device, which poses unique system challenges and learning opportunities, and paves the way for truly distributed learning  of privacy leakage.

\section{System Design \& Implementation} \label{sec:system}

\subsection{Goals and Design Rationale\label{sec:problem-leak}}

{\bf Problem Statement.} Mobile devices have access to a wealth of resources and information, much of which is personal and potentially sensitive. We will refer to such  personally identifiable information as {\em PII}. Examples include:
\begin{itemize}
\item Device Identifiers: IMEI, AndroidID, phone number, serial number, ICCID, MAC Address.
\item User identifiers: credentials (per app, usually transmitted over HTTPS), advertiserID, email.
\item User demographic:  first/last name, gender, zipcode, city, \etc\xspace - unavailable through Android APIs.
\item Location:  (latitude and longitude coordinates, available through Android APIs.
\item User-defined: the user can also define any custom string that should be monitored  (\eg see GUI in Fig. \ref{fig:guiPI}), such as digits of her credit card.
\end{itemize}

A key insight of  our design is the distinction on whether PII of interest is known to the device or not. We refer to PII that consists of strings known a priori on the device (\eg via Android APIs, or defined by the user) as \preDefLeak . We refer to PII that is not known to our \antshield system (\eg hidden from apps or changing dynamically) as  \unknownLeak. By default, we assume that any PII available via Android API calls are \preDefLeak (\eg IMEI, AndroidID, phone number, serial number, ICCID, MAC Address, advertiserID, email, and location), and the rest are \unknownLeak (\eg username login, password, first/last name, gender, zipcode, and city).

Our system employs different techniques to detect the transmission of \preDefLeak PII (\stringMatch) and \unknownLeak PII (classification).
We refer to the transmission of a packet from the device to the network,  containing at least one  PII, as a {\em privacy exposure (or leak)}. This transmission may be: (i) intended to collect information about the user; (ii)  benign, \eg necessary for the operation of the app, acceptable to the user, or (iii)  of the honest-but-curious nature. Distinguishing between privacy {\em exposure} and an actual privacy {\em leak} is out of the scope of this paper, and we refer to the two terms interchangeably, meaning ``exposure.'' Our goal is to detect privacy exposures
{\em on the device} with low overhead, accurately and in real-time. 
This is a first step towards enabling distributed learning of PII exposures.

{\bf Design Objectives and Choices.}  %
First, we want a solution that can be used by the non-sophisticated end-user: %
 a mobile app that the user can simply install (as an app, without rooting the phone), enable in the background, and occasionally interact with. %
Second, we want  a solution that operates purely on the device and does not redirect traffic through a middle server. This has several advantages: it does not need to expand the trust base (data does not need to leave the user's device), and it is well positioned to have access to rich information available on the device  (app names, and \preDefLeak PII). %
To meet both of these goals, we use the  \antLib v0.1.5 \cite{antmonitor-arxiv}. %
 To the best of our knowledge,   \anteater is the most efficient implementation (in terms of throughput, battery and other resources)  for on-device packet interception and inspection of both unencrypted and encrypted traffic, today; see \cite{antmonitor-arxiv} for details. \anteater relies on a VPN service on the device (but  not on a remote VPN server), which is the only way to intercept traffic today without rooting the phone.   If more efficient libraries for packet interception become available in the future, \antshield 's modular design allows to replace this component.

 Third, we want to accurately detect a comprehensive range of PII in real-time, both \preDefLeak (through \stringMatch), and the remaining \unknownLeak ones through machine learning models. Towards the first goal, we utilize the DPI  API provided by \anteater %
 \cite{antmonitor-arxiv}. Towards the second goal, we build specialized classifiers defined in  Sec. \ref{sec:system-implementation}.

\begin{figure}[t!]
	\centering
	\includegraphics[width=0.5\textwidth]{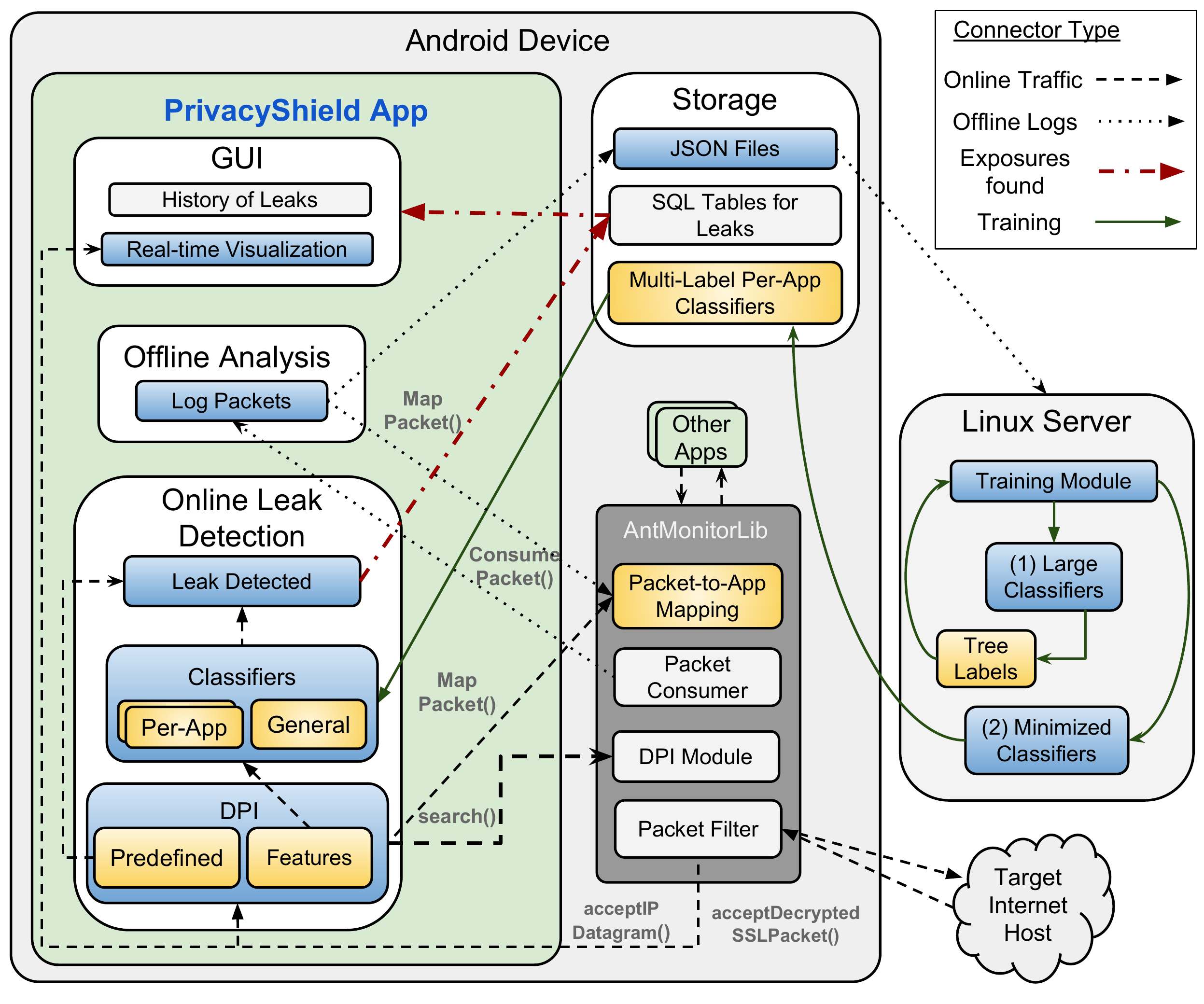}
	\caption{\footnotesize{\antShield Architecture. It consists of a mobile app on the device and a remote server. Real-time classification consists of the following steps: each packet is intercepted by \antLib, mapped to an app, and analyzed for multiple \preDefLeak and \unknownLeak leaks; 
	detection occurs before the packet is forwarded towards its remote destination (and an action may be taken to block the leak). Offline operations include loading and (re)training the classifiers, and (if the user agrees) uploading logs to the server.}}\label{fig:antShield}
\end{figure}

\subsection{PrivacyShield Architecture} \label{sec:system-antShield}

 The overview of the \antshield architecture is depicted in Fig. \ref{fig:antShield}. It consists of a mobile app and an (optional) server. A brief overview of each component %
 is described next.

 \begin{figure}[t!]
	\begin{center}
		\subfigure[\preDefLeak PII, possible actions, and custom filters (name)]{\hspace{6pt}\includegraphics[width=3.5cm]{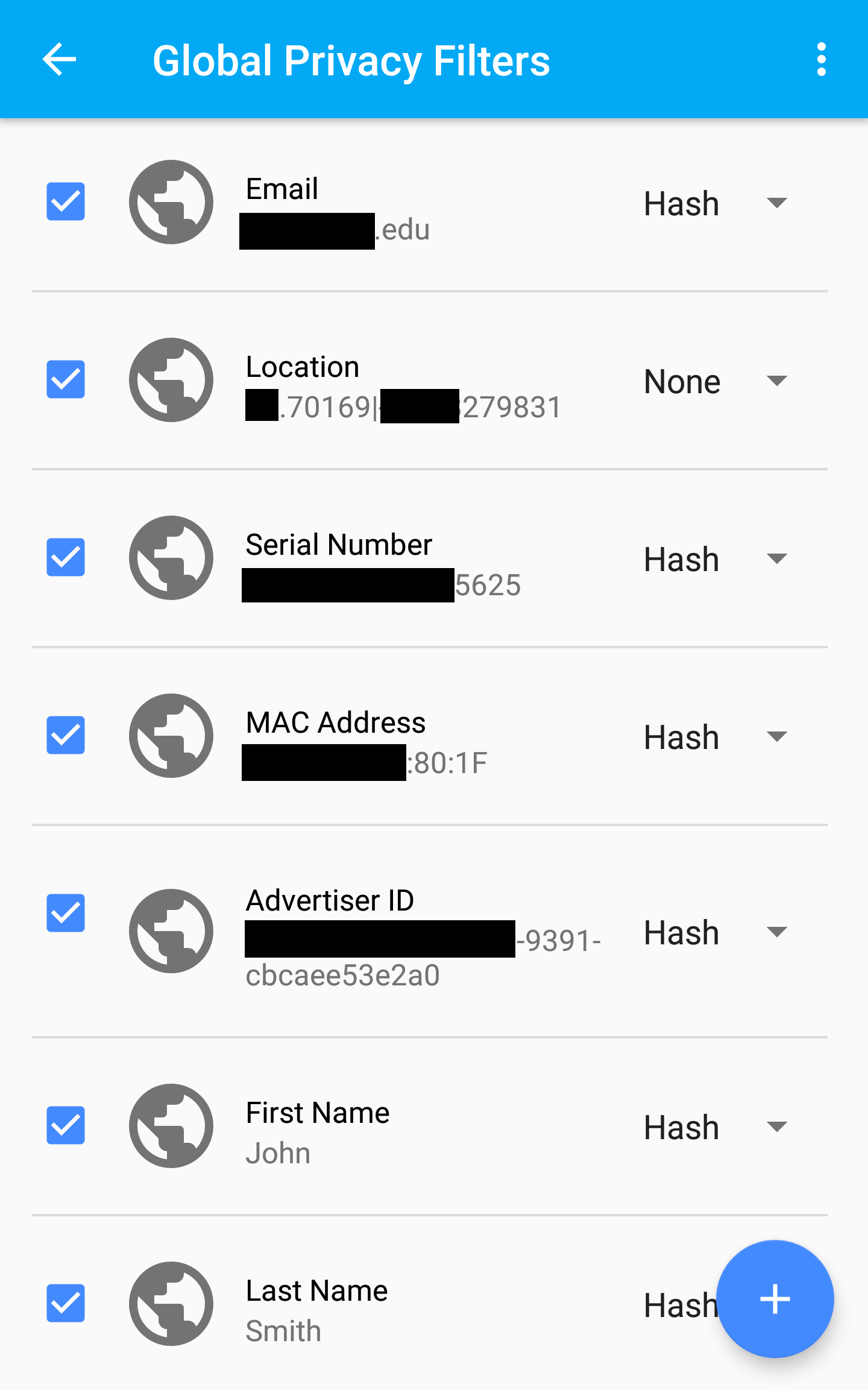}\label{fig:guiPI}\hspace{6pt}}
		\subfigure[Real-Time Visualization: which apps transmit PII to which remote servers]{\hspace{6pt}\includegraphics[width=3.5cm]{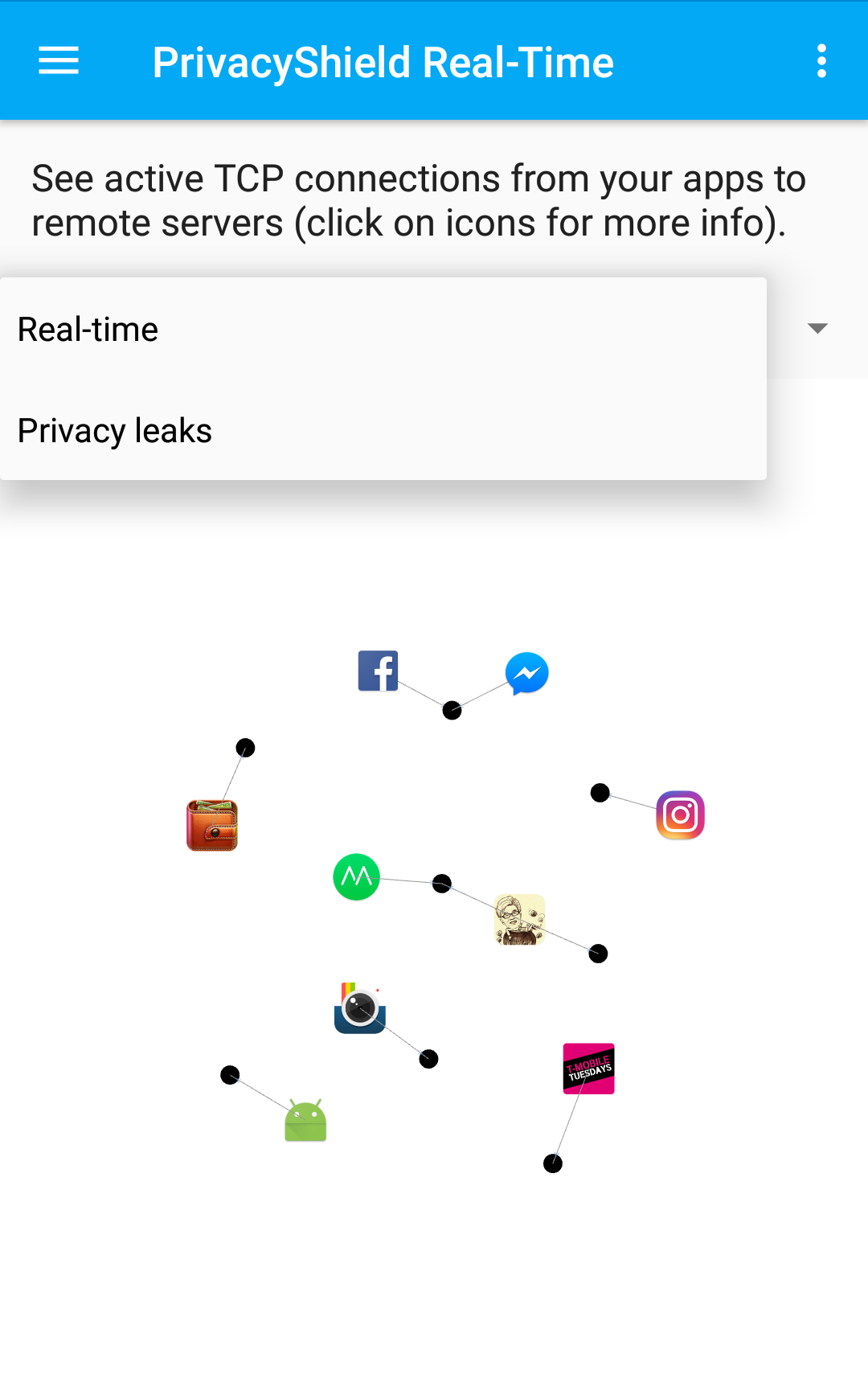}\label{fig:guiVisual}}
		\vspace{-10pt}
		\caption{Select Screenshots of the \antshield Android App.} 
		\label{fig:screenshots}
		\vspace{-15pt}
	\end{center}   
\end{figure}

{\bf Online Leak Detection.} %
This is the core  functionality of our  PII exposure detection. As shown in Fig. \ref{fig:antShield}, \antShield leverages calls to \antLib  ({\tt accept\-IP\-Datagram} and {\tt accept\-Decrypted\-SSL\-Packet})  
 to intercept packets (in clear text or decrypted SSL, respectively).
Each outgoing intercepted packet  is analyzed with DPI for \preDefLeak leaks and for features. The features are then passed to classifiers to detect \unknownLeak leaks (described in detail in Sec. \ref{sec:system-implementation}). Either way, if a PII exposure is detected, the user is notified and the exposure is logged. If the user chooses to, the leaky packet can be blocked, or it can be allowed to continue towards its remote destination. %

{\bf Offline Analysis.} This module can be used when heavier processing is required. %
 For example, to generate logs on the device, which require I/O operations, we use \antLib's {\tt con\-sumePacket()} API from this module. This module can be used to generate ground truth on the device;  and in the future, it can be extended to re-train classifiers on the device without sending data to a central server.

{\bf Storage (on-Device and/or Server).} The \antShield app comes pre-loaded with classifiers trained on our existing dataset (Sec. \ref{sec:data-ours}), so that users have no need to contact any server and can use the system as-is.  Only if the user chooses to  do so, logs (packet traces, JSON or other meta data) can be maintained on the device and/or occasionally be uploaded to a server. The use of the server is optional -- by default, logs do not need be collected or leave the device. The user may choose to share her data with the server to get the benefits of crowdsourcing, and retrained classifiers. As an example scenario, we used the logging capability of the Offline Analysis module to generate the dataset used in our evaluation. Specifically, each captured packet was labeled with the type of PII that it was leaking and the app name that generated the packet. The PII itself was replaced, and the packet was converted to a JSON format for easier processing at the machine learning training stage (see Sec. \ref{sec:data-ours} for details). In general, this feature is useful for other researchers  who may wish to generate their own datasets manually or from user studies.

{\bf GUI.} This component has two purposes. It allows the user to specify various preferences,
 most importantly, the \preDefLeak PII to be monitored. By default, these include PII available on the device through Android APIs, as shown in Fig.  \ref{fig:guiPI}.  Users that trust \antShield can also opt-in and predefine additional PII, such as  name and gender or any string (\eg digits of a credit card). Second,  \antshield's GUI notifies the user about PII exposed. From here, users can decide to allow the leak to happen, replace the exposed PII with a random string of the same length (so as not to alter the payload size), or block the packet completely. Whatever action the user selects, it is remembered for future occurrences of the same PII/app combination. Users can view a history of leaks at any time and can also see where each app sends data as a graph of connections updated in real-time (Fig. \ref{fig:guiVisual}). The edges of the graph can be filtered or annotated by the leaked PII, and more information about the remote servers receiving the leak can be displayed.

\subsection{Leak Detection Methodology}\label{sec:system-implementation}

At the heart of \antShield lies the online inspection of network packets to detect if they contain PII.

First, we use a {\em hybrid} \stringMatch-classification methodology. As described in Sec. \ref{sec:problem-leak}, a key insight is that PII can be split  into two categories: \preDefLeak and \unknownLeak, depending on whether  they are known a priori or not. This is an inherent advantage of operating on the device: \antShield has access to all the \preDefLeak strings and can use DPI to search for them; we refer to this method as \stringMatch. This not only gives us 100\% accuracy on finding \preDefLeak leaks, if they are not obfuscated, but also reduces the set of PII that classifiers must learn, thereby improving the accuracy of finding \unknownLeak leaks and reducing variance (see Sec. \ref{sec:prediction-result}).

Second, we treat PII detection as a \multiLabel problem, since a packet may contain zero, one, or multiple PII. Our classifiers decide,  in one step, if any PII are contained in a packet, and if so - what type. More specifically, we use Mulan \cite{mulan} %
 to perform multi-label classification using the Binary Relevance (BR) transformation method \cite{tsoumakas2006multi}. The idea is to train a separate binary classifier for each label. Since the C4.5 DTs worked well for classifying leak vs non-leak, we use them as the independent classifiers in BR. %

Third, we build classifiers {\em per-app}, instead of {\em per destination domain}. This is possible thanks to   \antshield running on the device:  it can accurately map a packet to the app that generated it. %
From a classification point of view, per-app classifiers perform similarly to per-domain classifiers, as shown in Sec. \ref{sec:prediction-result} and explained in Sec. \ref{sec:findings}. However,  per-app classifiers have important system advantages. First, they allow for easy setup and scalability: only the few classifiers for the installed apps on the particular device must be loaded into memory. This is much smaller than hundreds of domains contacted by those apps and the third-party libraries contained within them. Second, they apply to all TCP and UDP traffic, not just to HTTP(S) traffic. Third,  per-app classifiers obviate the need for DNS lookups, which are costly and inaccurate, but are necessary when using per-domain classifiers. \recon parsed HTTP(S) packets to extract the host name (which is also costly in terms of CPU) and decided which per-domain classifier to apply. One possible solution is to do reverse-DNS lookup to map (all TCP and UDP, not only HTTP(S)) packets to their intended hosts. However, many companies opt-in to use third-party web service providers (such as Amazon AWS), and for them, reverse-DNS returns host names that are not very useful (\eg ec2-54-164-159-29.compute-1.amazonaws.com). As a \workaround, it may be possible to implement a reverse-DNS cache on the device by keeping track of all the DNS requests. Unfortunately, we have seen many cases where the same IP maps to multiple host names (again, due to third-party web service usage). Finally, even if we could somehow achieve perfect mapping of IPs to host names, there is still the problem of domain name extraction. One solution is to maintain a public suffix list, but that would take up too much memory on the mobile device. Another solution is to keep removing prefixes from the host name and do DNS queries until a Start of Authority record is reached; but this would cause too much network delay on each packet before it can even be assigned a classifier.

\subsection{Real-time Implementation on the Mobile Device} \label{sec:opt}
The classifiers described in the previous section have value on their own right. However, it is highly non-trivial to apply them in real-time on a mobile device, with limited CPU and RAM. \antshield is the first system to achieve this goal 
thanks to the following system optimizations.

{\bf Detecting PII in an Outgoing Packet.}
Our hybrid approach relies on  \stringMatch to search for the \preDefLeak leaks and on classification methodology to detect \unknownLeak ones. The former benefits from the good performance of \antLib's efficient DPI module. The latter needs to parse packets to extract words that are used as features of the classifiers.  With off-the-shelf \recon, to extract words from a packet,  several invocations of Java string parsing methods would be required, which  are extremely slow on a mobile device. We were able to  extract features from the traffic while completely avoiding parsing by exploiting the following observation:  most decision trees are one-level deep and only a third of the trees have a depth greater than two. 
Therefore, we only need to extract the words that appear in the decision tree nodes and we can use DPI to search for them. Since the Aho-Corasick algorithm used in the \antLib can search for many strings in one pass of the packet, having these extra words to  search for does not affect performance. 

Extracting words that appear in the decision tree nodes and using DPI to search for them works well in most cases. However, in some cases the words are too small and can actually be part of a longer word. In this case, DPI search would mark a feature as existent, when in fact it's part of a different word, causing an incorrect prediction. As an example, {\em hulu} was receiving the word `\textvtt{profile}' in the packets that also contained the user's first name. However, many packets that did not contain any exposures, contained the word `\textvtt{video\_profile}.' To avoid these DPI-based false positives, we decided to keep the delimiters surrounding each word during feature selection. So, in the case of {\em hulu}, we used `\textvtt{/profile?}' as the feature. This trick allowed us to extract the same words with DPI as with Java parsing.

{\bf Minimizing Classifiers to Load in RAM.} 
With limited RAM, care must be taken when loading machine learning models from disk to memory. To minimize the impact on RAM, we: (i) load per-app models  only for those apps that are installed on the device; and (ii) perform a two-step training method to reduce the general classifier feature set (see Fig. \ref{fig:antShield}). Specifically, the general classifier has a feature set size of over 12k, and during prediction needs the  allocation of a double array with size 12k+. While this is a small size for a server, on the mobile device it causes major issues -- if one loads the full general classifier, most web pages and applications do not load. This is because each time a packet has to be predicted by the general classifier (when there is no corresponding per-app classifier), the memory allocation becomes so large that a {\em blocking garbage collection call} has to be executed by the Android OS after every prediction. This blocks the main networking thread, causes connections to time-out, and prevents pages from loading.
We were able to reduce the feature set by exploiting the existing classifier tree: we re-trained the general classifier using only the words that appear in the tree nodes as features. This resulted in a feature set size of only 509, \ie  a 24x reduction for the %
general classifier, which in itself allowed \antShield  to run in real time. %
 Further improvements were achieved by reducing the feature set of per-app classifiers. %
Overall, \antShield's memory usage was around 100 MB, which is acceptable: many  popular apps, \eg\xspace  {\em Facebook}, use as much as 200 MB RAM.

{\bf Real-Time Packet-to-App Mapping.}
In order to call the per-app classifiers, we first need to map a packet to the application that generated it.
The \antLib provides packet-to-app mapping but only off-line (\eg after a packet has been read off a queue on a different thread that does not block the main networking thread) \cite{antmonitor-arxiv}, which is not fast enough to run on-line. Specifically, when using \antLib's original mapping implementation, we were only able to reach a throughput of 1 Mbps when testing with {\em Speedtest}. Upon further code and CPU usage analysis, we found that the inefficiency stemmed from two issues: (i) the \antLib was doing some Java string parsing to extract the app UID, source IP/port, and destination IP/port that were separated by a comma when returned from the native C module; (ii) the \antLib was storing the mappings in a HashMap keyed by a String (made of concatenating source/ destination IP/port numbers), which caused many String comparisons whenever an item needed to be fetched from the HashMap. To avoid these costly operations, we changed \antLib's native C module to return the app UID and the source port number only, as separate elements in an array. (it is best to avoid using complex data structures in native C). This way the Java part of the code could separate out the UID (and fetch the corresponding app name) and the source port of each open connection without doing any parsing. The source port number is then used as the key to the HashMap that fetches the corresponding app names. These improvements allowed us to do real-time packet-to-app mapping while achieving network speeds close to regular device operation speeds.

{\bf Real-time.} The evaluation in Sec. \ref{sec:eval-time}  shows that our optimizations make the crucial difference for  being able to detect PII in real-time on the device: 1ms for extracting words (as opposed to 30ms+ if parsing out all words) and 1ms for classification.%

\section{Evaluation} \label{sec:eval}

\subsection{PrivacyShield Datasets} \label{sec:data-ours}

\begin{center}
	\begin{table}[t!]
		{\scriptsize
			\begin{tabular}[b]{|c|p{0.8cm} p{0.8cm}|p{0.8cm} p{0.8cm}|} 
				\hline
				&\textbf{ReCon Public dataset(s)}  &&   \textbf{\antshield dataset(s)} &\\[0.5ex]
				\hline\hline
				&\textbf{Auto} &  \textbf{Manual} & \textbf{Auto} &  \textbf{Manual}\\[0.5ex] 
				\hline
				\# of Apps & 564 & 91 &414 & 149\\
				\hline
				\# of packets & 16761 &13079 &  21887 & 25189\\
				\hline
				\# of destination domains & 450 &368 & 597 & 379\\
				\hline
				\# of leaks detected & 1566 & 1755  & 4760 &3819\\
				\hline
				{\bf \# of \unknownLeak leaks} & 4 &78 & 483 & 516\\
				\hline
				\# of leaks in encrypted traffic & - & - & 1513 & 1526\\
				\hline
				\# of packets with {\bf multiple leaks} &50  & 224  & 1506 & 790\\
				\hline
				
				{\bf \# of background leaks} & - & - & 2289 & 639\\
				\hline
				\# of HTTP packets & 16761 & 13079 & 13694 &13648\\ 
				\hline
				\# of HTTPS packets & - & -  &6830 &8103\\
				\hline
				{\bf \# of TCP packets} & - & - &867 &2264\\
				\hline
				{\bf \# of leaks in TCP (other ports)} & - & - & 38 &7\\
				\hline
				{\bf \# of UDP packets} & - & - & 496 &1174\\  
				\hline
				{\bf \# of leaks in UDP} & - & -  & 17 &12\\ [1ex] 
				\hline
			\end{tabular}
		}
		\caption{Summary of Datasets.  \recon is the previous state-of-the-art, collected in the middle of the network \cite{recon15}. \antShield's  Manual and Automated datasets were collected on the device.}
		\label{tbl:summary}	
	\end{table}
\end{center}\mbox{}

\vspace{-35pt}

In order to evaluate the effectiveness of our methodology in detecting  private information exposure, we collected and analyzed two \antShield Datasets. 
We logged all packets generated by different apps on a test device (Nexus 6) and converted each packet into a JSON object that reported any PII exposures (see \ref{sec:problem-leak} for a list) and broke the packet into any relevant fields (destination IP address/port, HTTP method, if applicable, and etc).
We collected  two different datasets, depending on how we interacted with apps, described next.

{\bf Manual Testing.}  First, in order to assess PII leaks during typical user behavior, 
we tested 100 most popular and free Android  apps, based on rankings in {\em AppAnnie} \cite{appannie}. %
We tested in batches:  we installed 5 apps on the test device %
and then used \antShield to intercept and log packets while interacting with each app for 5min. 
After all apps in the batch were tested, we switched off the screen and waited 5min to catch  any packets  sent in the background. Next, we uninstalled each app %
and finally, turned off \antShield. 

{\bf Automatic Testing.} %
We also used  the {\em UI/Application Exerciser Monkey} \cite{monkey} to automatically interact with apps.  This does not capture typical user behavior but enables extensive and stress testing of more apps. We  installed 4 batches of 100 applications each, and had {\em Monkey} perform 1,000 random actions in each tested app
while \antShield logged the generated traffic. 
At the end of each batch, we switched off the screen of the test device and waited for 10min  to catch additional exposures sent in the background. %

{\bf Summary.}  Since the two (Automatic and Manual)  \antShield Datasets  capture different behaviors, we describe and analyze them separately.  However, for the purposes of training and testing classifiers, we merged them into one, referred to as the \antshield Dataset. %
The  \antShield datasets are  summarized in Table \ref{tbl:summary}, next to the prior state-of-the-art PII datasets collected by \recon  \cite{recon15}. %

Using \antShield to capture packets on the device has several advantages compared to  previous datasets collected in the middle of the network: (1) we were able to accurately map each packet to the app that  generated it; (2) we kept track of foreground vs. background apps, to see what kind of data apps send while in the background; (3) we gained insight into TLS, UDP, and regular TCP traffic, in addition to HTTP; (4) scrubbing PII and labeling packets with the type of PII they leak was fully automated: \antShield already provides \preDefLeak strings, and we entered the \unknownLeak strings (\eg fake test account credentials) as custom filters (as in Fig. \ref{fig:guiPI}).
The resulting dataset contains more %
and  richer %
information about exposures than before. Some  advantages are inherent to running on the device (\ie the ability to capture contextual information, including the app names). Other differences are due to changes in app versions and leak behavior over time. Therefore, in addition to being used to evaluate our  methodology (Section \ref{sec:prediction-result}), our datasets have value on their own and we will make them available to the community.

\subsection{Exposures Found in the Datasets} \label{sec:findings}

Our datasets provide us with insights into the current state of privacy leaks in the Android ecosystem. Some of the captured patterns were previously unknown, and are revealed for the first time here. 
For example, we were able to detect leaks happening in the background, leaks in plain TCP and UDP (not belonging to  HTTP(S) flows), two orders of magnitude more  \unknownLeak leaks than before (which is crucial for training classifiers), several hundreds of packets with not one but multiple leaks  (which motivated our  \multiLabel approach), and malicious scanning for rooted devices.

\begin{table*}[t!]
	\centering
	{\scriptsize
		\begin{tabular}{|@{}l@{}|@{}l@{}|@{}l@{}|}
			\hline
			{\bf \xspace App Name} & {\bf \xspace  Leak Type}  &  {\bf \xspace  \# Leaks \:}\\[0.5ex] 
			\hline
			\xspace com.roblox.client 2.280.107211 & \xspace Username, Location  &  \xspace 1234 \\
			
			\xspace com.ss.android.article.master3.2.7 & \xspace  \multirow{2}{*}{  \parbox{3.5cm}{City, Adid, Location, AndroidId, IMEI}}  & \xspace 766\\
			& & \\
			
			\xspace com.cleanmaster.security3.2.6  & \xspace Adid, AndroidId  & \xspace 511 \\
			
			\xspace com.cyberpony.stickman.warriors.epic1.3  & \xspace Adid, City, Location, Zipcode & \xspace 434 \\
			
			\xspace com.paypal.android.p2pmobile6.9.0 & \xspace \multirow{2}{*}{\parbox{3.5cm} {City, FirstName, LastName, SerialNumber, Zipcode, Adid, AndroidId, Password, Email}}  &\xspace 257 \\
			& & \\
			& & \\
			
		    \xspace com.weather.Weather7.7.1  &\xspace Adid, Location &\xspace 172 \\
		    
			\xspace com.pof.android3.45.2.1417399  &  \xspace Adid, Username, AndroidId &\xspace 136 \\
			
			\xspace com.bitmango.go.wordcookies1.1.9  & \xspace Adid, AndroidId  &\xspace 135 \\
			
			\xspace com.kiloo.subwaysurf1.68.0  & \xspace Adid, IMEI, AndroidId  &\xspace 101\\
			
			\xspace com.qisiemoji.inputmethod5.5.8.1570  &\xspace Adid, IMEI, AndroidId  & \xspace99 \\
			
			\xspace com.bitmango.go.blockhexapuzzle1.3.7  & \xspace Adid, AndroidId  &\xspace 97 \\
			
			\xspace com.bitmango.rolltheballunrollme1.6.5 & \xspace Adid, AndroidId  &\xspace 95 \\
			
			\xspace com.jb.zcamera2.48  &\xspace  \multirow{2}{*}{\parbox{3.5cm}{ Adid, AndroidId, IMEI, Email, IMSI}}  &\xspace 87 \\
			 & & \\

			\xspace $\cdots$ & \xspace $\cdots$ & \xspace $\cdots$ \\
			\hline
			\xspace All & \xspace All & \xspace 3819 \\[0.5ex] 
			\hline
		\end{tabular}
		\quad \quad
		\begin{tabular}{|@{}l@{}|@{}l@{}|@{}l@{}|}
			\hline
			{\bf\xspace Domain Name} &  {\bf \xspace Leak Type}  & {\bf\xspace \# Leaks \:}\\[0.5ex] 
			\hline
			\xspace isnssdk.com & \xspace Adid, IMEI. AndroidId  &\xspace 739 \\
			\xspace roblox.com & \xspace Location &\xspace 679\\
			\xspace facebook.com & \xspace Adid  & \xspace651 \\
			\xspace rbxcdn.com  & \xspace Location  &\xspace 561 \\
			\xspace mopub.com & \xspace Adid  &\xspace 340 \\
			\xspace bitmango.com  &\xspace Adid &\xspace 262 \\
			\xspace paypal.com &  \xspace AndroidId &\xspace 257 \\
			\xspace appsflyer.com  & \xspace Adid, AndroidId  &\xspace 239 \\
			\xspace goforandroid.com  &\xspace Adid, IMEI, IMSI, AndroidId & \xspace 171 \\
			\xspace applovin.com & \xspace Adid & \xspace 157\\
			\xspace pof.com  & \xspace Adid, AndroidId  & \xspace 121 \\
			\xspace adjust.com & \xspace Adid  & \xspace 96 \\
			\xspace adkmob.com  &\xspace Adid, AndroidId  &\xspace 88 \\
			\xspace pandora.com   & \xspace Adid, AndroidId, Zipcode  & \xspace 78 \\
			\xspace wish.com  & \xspace Adid  &\xspace 78 \\		
			\xspace lyft.com & \xspace Location  &\xspace 68\\			  			  
			\xspace $\cdots$ & \xspace $\cdots$ & \xspace $\cdots$ \\
			\hline
			\xspace All & \xspace All & \xspace 3819 \\[0.5ex] 
			\hline
		\end{tabular}
	}
	\caption{Manual dataset: Summary of applications and domain names with most leaks and their leak types}
	\label{tab:leaks}
\end{table*}

\begin{table*}[t!]
	\centering
	{\scriptsize
		\begin{tabular}{|@{}l@{}|@{}l@{}|@{}l@{}|}
			\hline
			{\bf\xspace App Name} &  {\bf \xspace Leak Type}  & {\bf \xspace \# Leaks \: }\\[0.5ex] 
			\hline
			\xspace cmbinc12.mb32b5.98 &\xspace  \multirow{2}{*}{  \parbox{3.5cm}{City, Adid, Location, AndroidId, Zipcode}}  & \xspace 1326 \\
			& & \\
			\xspace com.kitkatandroid.keyboard3.9.9 &\xspace  \multirow{2}{*}{  \parbox{3.5cm}{Adid, , Location, AndroidId, SerialNumber}}  & \xspace 1046\\
			& & \\
			\xspace com.episodeinteractive.android.catalog5.61.1+g  &\xspace \multirow{2}{*}{  \parbox{3.5cm}{ \xspace Adid, Gender, SerialNumber, AndroidId}}&\xspace 438 \\
			& & \\
			\xspace com.myyearbook.m11.8.0.681 & \xspace Adid, City, Location, Zipcode &\xspace 263 \\
			\xspace System0.1.5 &\xspace \multirow{2}{*}{\parbox{3.5cm}{ Username, City, Zipcode, Adid, AndroidId, Location, IMEI, IMSI}}  & \xspace 255 \\
			& & \\
			\xspace com.clearchannel.iheartradio.controller7.2.2  &\xspace Adid, Zipcode & 220 \\
			\xspace com.cmcm.live3.4.9  &\xspace \multirow{2}{*}{\parbox{3.5cm}{Adid, AndroidId, Location, IMEI, SerialNumber, IMSI}}  &\xspace 213 \\
			& & \\
			\xspace com.apalon.myclockfree2.29 &\xspace Adid, City &\xspace 206 \\
			\xspace com.freecraft.pocket.edition2.0  &\xspace Adid, City, Location, Zipcode  &\xspace 174 \\
			\xspace com.madebyappolis.spinrilla2.2.4  &\xspace \multirow{2}{*}{\parbox{3.5cm}{Adid, City, Location, AndroidId, Zipcode}} &\xspace 146\\
			& & \\
			\xspace $\cdots$ &\xspace $\cdots$ &\xspace $\cdots$ \\
			\hline
			\xspace All & \xspace All & \xspace 4760 \\[0.5ex] 
			\hline
		\end{tabular}
		\quad \quad
		\begin{tabular}{|@{}l@{}|@{}l@{}|@{}l@{}|}
			\hline
			{\bf \xspace Domain Name} &  {\bf \xspace Leak Type}  & {\bf \xspace \# Leaks \: }\\[0.5ex] 
			\hline
			\xspace mopub.com & \xspace Adid  &\xspace 2878 \\
			\xspace pocketgems.com & \xspace AndroidId &\xspace 416\\
			\xspace applovin.com & \xspace Adid  &\xspace 409 \\
			\xspace appsflyer.com  & \xspace Adid  &\xspace 312 \\
			\xspace ksmobile.net & \xspace SerialNumber, Location, AndroidId  &\xspace 216 \\
			\xspace ihrhls.com  &\xspace Adid &\xspace 215 \\
			\xspace goforandroid.com &  \xspace AndroidId &\xspace 210 \\
			\xspace tapjoyads.com  &\xspace IMEI, AndroidId  &\xspace 169 \\
			\xspace amplitude.com & \xspace Adid &\xspace 152\\
			\xspace lkqd.net  & \xspace Adid, AndroidId  &\xspace 150 \\
			\xspace tapjoy.com & \xspace Adid, AndroidId  &\xspace 148 \\
			\xspace pinterest.com  &\xspace AndroidId  &\xspace 112 \\
			\xspace bitmango.com   & \xspace Adid  &\xspace 109 \\
			\xspace 3g.cn  & \xspace AndroidId  &\xspace 107 \\		
			\xspace instagram.com& \xspace Username  &\xspace 105\\	
			\xspace crashlytics.com  & \xspace Adid, AndroidId  &\xspace 98 \\		  			  
			\xspace $\cdots$ &\xspace $\cdots$ &\xspace $\cdots$ \\
			\hline
			\xspace All &\xspace All &\xspace 4760 \\[0.5ex] 
			\hline
		\end{tabular}
	}
	\caption{Auto dataset: Summary of applications and domain names with most leaks and their leak types}
	\label{tab:leaks2}
\end{table*}

\begin{table}[t!]
	\centering
	{\small
		\begin{tabular}{|p{4.8cm}|p{1.6cm}|p{0.8cm}|}
			\hline
			{\bf App Name} &  {\bf Leak Types}  & {\bf Port}\\[0.5ex] 
			\hline
			\xspace System  &\xspace IMEI, IMSI, AndroidId &\xspace 8080 \\
			\xspace com.jb.gosms &\xspace  AndroidId  &\xspace 10086 \\
			\xspace com.jiubang.go.music &\xspace AndroidId &\xspace 10086 \\
			\xspace air.com.hypah.io.slither &\xspace Username  &\xspace 10086 \\
			\xspace com.jb.emoji.gokeyboard &\xspace AndroidId &\xspace 10086 \\
			\xspace com.gau.go.launcherex &\xspace AndroidId  &\xspace 10086 \\
			\xspace com.steam.photoeditor &\xspace AndroidId &\xspace 10086 \\
			\xspace com.jb.zcamera &\xspace AndroidId  &\xspace 10086\\
			\xspace com.flashlight.\-brightestflashlightpro &\xspace AndroidId &\xspace 10086\\[0.5ex] 
			\hline
		\end{tabular}
		
		\quad \quad \quad
		
		\vspace{5pt}
		
		\begin{tabular}{|@{}l@{}|@{}l@{}|@{}l@{}|}
			\hline
			{\bf Domain Name} &  {\bf Leak Types}  & {\bf Port}\\[0.5ex] 
			\hline
			\xspace 206.191.155.105 &\xspace Username  &\xspace 454 \\
			\xspace 206.191.154.41 &\xspace Username &\xspace 454 \\
			\xspace 23.236.120.208 &\xspace AndroidId  &\xspace 10086 \\
			\xspace 3g.cn  &\xspace IMEI, IMSI, AndroidId &\xspace 8080 \\
			\xspace 23.236.120.220 &\xspace AndroidId & \xspace10086 \\[0.5ex] 
			\hline
		\end{tabular}
	}
	\caption{TCP packets (non HTTP/S) sending PII over ports other than 80, 443, 53}
	\label{tab:tcp_leaks}
\end{table}

\begin{figure*}[t!]
	\begin{center}
		\centering
		\includegraphics[width=0.95\textwidth]{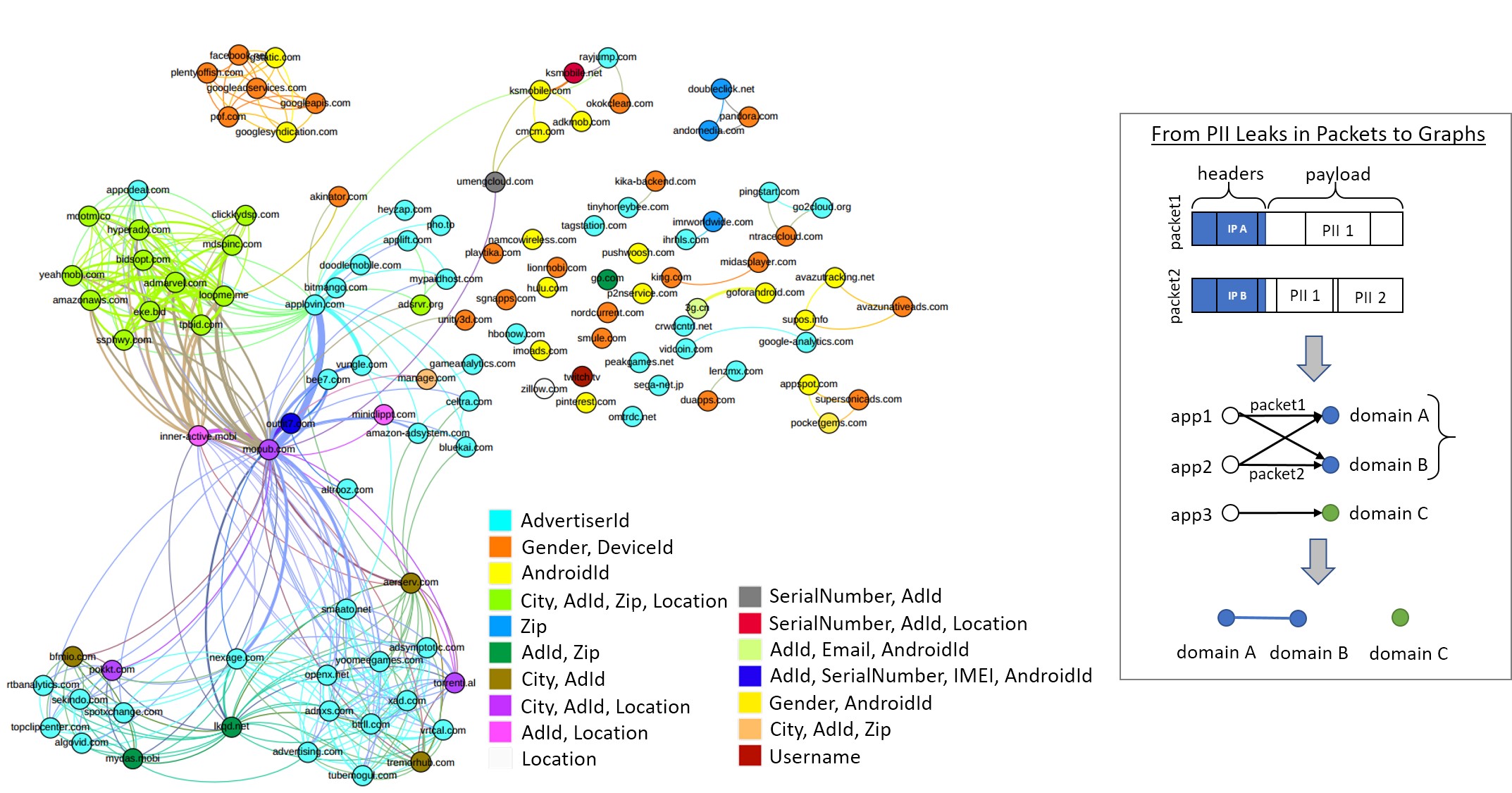}
		\caption{{\bf Understanding the behavior of leaking through graph analysis of the \antShield dataset.} The graph consists of nodes corresponding to destination domains and edges  representing the similarity of two domains. Two domains are similar if there are common apps that send packets with PII exposures to both domains; the more common apps  leak to these domains, the more similar they are, the larger the width of an edge between them. The color of a domain node indicates the types of  PII it receives. One can observe from the graph structure that domains form communities that capture interesting patterns: (1) The large communities on the left and bottom consist mostly of ad networks; ad exchanges are nodes in between ad communities; (2) Facebook/Google domains are a different community on their own, on the top left; (3)  small apps contact only their own domain, leading to isolate domain nodes; (4) domains in the same community receive the same type of leaks (as indicated by the color of nodes). }
		\label{fig:graph}
		\vspace{-15pt}
	\end{center}   
\end{figure*}

{\bf Background Leaks}. \antShield is in a unique position to capture leaks that happen in the background vs. foreground, and other contextual information that is only available on the device. Table \ref{tbl:summary} shows that there is a substantial number of background leaks (\eg half of all leaks in the automatic dataset) that should be brought to users' attention and be incorporated into learning algorithms.
We observed an order of magnitude more background leaks in the top apps in the  automatic vs the manual datasets. One possible explanation is that the random clicks in the automated test lead to clicking on ads, which generate traffic even after the app moves to the background.

{\bf Auto vs. Manual}. Tables \ref{tab:leaks} and \ref{tab:leaks2} show the top apps/domains that collect the most PII in our manual and auto datasets, respectively. We find that the top apps and domains differ for each dataset, indicating that it is important to test apps manually so as to fairly represent what happens to real users. For instance, we see that the auto dataset is more ad-oriented, while the top domains for the manual dataset include non-ad networks such as \textvtt{facebook}, \textvtt{paypal}, and \textvtt{pandora}. This is most likely due to the fact that our {\em Monkey} tests ended up clicking on ads during the random events, whereas real users tend to avoid ads.

{\bf Non-HTTP Leaks}.  Prior state-of-the-art datasets \cite{recon15}  reported only HTTP(S) leaks. Table \ref{tbl:summary} reports, for the first time, leaks in non-HTTP(S), including plain TCP or UDP packets. Our dataset contains 29 UDP leaks, all of which were exposing Advertiser Id and Location. As shown in Table \ref{tab:tcp_leaks}, we also found some apps (mostly games and photo-editing apps) that leaked the device ID over non-standard (80, 443, 53) TCP ports, such as 8080 or 10086 (a port known to be used  by trojans, Syphillis and other threats \cite{portsguide}). The destination IPs could not be resolved by DNS, indicating that the application may have hard-coded those IPs. We were also able to detect ver 3000 TCP packets with PII exposure, most of which are TCP segments, belonging to a larger HTTP packet. It is important to be able to classify these packets as well, since we will be receiving them through the VPN during real time inspection in \antshield. 

\begin{table*}[t!]
	\centering
	{\scriptsize
		\begin{tabular}{|@{}l@{}|@{}l@{}|@{}l@{}|}
			\hline
			{\bf \xspace App Name} &  {\bf \xspace Leak Type}  & {\bf \xspace \# Leaks \:}\\
			\hline
			\xspace com.ss.android.article.master3.2.7 &\xspace  \multirow{2}{*}{  \parbox{3.5cm}{City, Adid, Location, AndroidId, IMEI}}  &\xspace 752 \\
			& & \\
			\xspace com.cleanmaster.security3.2.6 & \xspace Adid, AndroidId  &\xspace 174\\
			\xspace com.paypal.android.p2pmobile6.9.0  &\xspace \multirow{2}{*}{  \parbox{3.5cm}{ City, FirstName, LastName, Zipcode, Adid, SerialNumber, AndroidId, Password, Email}}&\xspace 131 \\
			& & \\
			& & \\
			\xspace com.offerup2.3.12&\xspace \multirow{2}{*}{\parbox{3.5cm} {Adid, Username, FirstName, Location, Zipcode, AndroidId} }  &\xspace 114 \\
			& & \\
			\xspace com.cmcm.live3.4.9 &\xspace \multirow{2}{*}{\parbox{3.5cm}{Adid, AndroidId, Location, IMEI, SerialNumber, IMSI}}  &\xspace 114 \\
			& & \\
			\xspace me.lyft.android4.20.3.1439781 &\xspace  \multirow{2}{*}{\parbox{3.5cm}{City, FirstName, LastName, SerialNumber, Zipcode, PhoneNumber, Location, AndroidId}}   &\xspace 112 \\
			& & \\
			& & \\
			\xspace com.pinterest5.6.2  &\xspace Adid, AndroidId &\xspace 111 \\
			\xspace com.weather.Weather7.7.1  & \xspace Adid, Location &\xspace 110 \\
			\xspace com.qisiemoji.inputmethod5.5.8.15709 &\xspace Adid, IMEI, AndroidId &\xspace 83 \\
			& & \\		  			  
			\xspace $\cdots$ &\xspace $\cdots$ &\xspace $\cdots$ \\
			\hline
			\xspace All &\xspace All &\xspace 3039 \\
			\hline
		\end{tabular}
		\quad \quad
		\begin{tabular}{|@{}l@{}|@{}l@{}|@{}l@{}|}
			\hline
			{\bf\xspace Domain Name} &  {\bf \xspace Leak Type}  & {\bf\xspace \# Leaks \:}\\
			\hline
			\xspace mopub.com & \xspace Adid  &\xspace 2380 \\
			\xspace isnssdk & \xspace AndroidId, IMEI & \xspace805\\
			\xspace roblox.com  & \xspace Location  &\xspace 679 \\
			\xspace applovin.com & \xspace Adid  &\xspace 566 \\
			\xspace rbxcdn.com & \xspace Location  &\xspace 561 \\
			\xspace appsflyer.com  &\xspace Adid  &\xspace 549 \\
			\xspace facebook.com  & \xspace Adid  &\xspace 391 \\
			\xspace bitmango.com   & \xspace Adid  &\xspace 371 \\	
			\xspace goforandroid.com &  \xspace AndroidId &\xspace 262 \\	
			\xspace ihrhls.com  &\xspace Adid &\xspace 219 \\
			\xspace pocketgems.com  & \xspace AndroidId  &\xspace 211 \\
			\xspace ksmobile.net & \xspace SerialNumber, Location, AndroidId  &\xspace 159 \\
			\xspace tapjoy.com & \xspace Adid, AndroidId  &\xspace 151 \\
			\xspace tapjoyads.com  &\xspace IMEI, AndroidId  & \xspace147 \\
			\xspace wish.com & \xspace Adid &\xspace 139\\
			\xspace paypal.com  & \xspace AndroidId  &\xspace 131 \\		
			\xspace pof.com& \xspace pof.com  &\xspace 122\\	
			
			\xspace $\cdots$ & \xspace $\cdots$ & \xspace $\cdots$ \\
			\hline
			\xspace All &\xspace All &\xspace 3039 \\
			\hline
		\end{tabular}
	}
	\caption{{\small Summary of applications and domain names with HTTPS leaks in our dataset (manual and auto)}}
	\label{tab:leaks_https}
\end{table*}

{\bf HTTPS Leaks}. Since traffic is increasingly over HTTPS than HTTP, we need to inspect and train on HTTPS traffic as well. However, due to their sensitive nature,  previous HTTPS datasets \cite{recon15} were not made publicly available and we had to collect our own. Table \ref{tab:leaks_https} summarizes the leaks we discovered in HTTPS traffic. The top app \textvtt{com.ss.android.\-article.\-master} is a news app, thus it makes sense for it to query the user's city, perhaps to fetch localized news. However, it is unclear why the app needs the user's IMEI (when it already has the AdId) and the specific longitude and latitude coordinates of the user. Another example is \textvtt{com.cmcm.live} - it leaks 5 different device identifiers for no apparent reason. Hence, although well-behaving apps should use HTTPS, they should also be inspected for potential privacy leaks as not all information that they gather is necessary for their functionality. We also found that the majority of top domains receiving PII over HTTPS were ad-related.

\begin{table}[t!]
	\centering
	{\scriptsize
		\begin{tabular}{|p{2.5cm}|p{2cm}|p{2.2cm}|}
			\hline
			{\bf\xspace App Name} &  {\bf\xspace Domain}  & {\bf\xspace Leak Types}\\
			\hline
			\xspace com.bitstrips.imoji 10.2.32, 10.3.76 &\xspace pushwoosh.com & AndroidId \\
			\xspace com.nianticlabs\-.pokemongo 0.57.4 &\xspace upsight-api.com  & Location, AndroidId \\
			\xspace com.psafe.msuite 3.11.6 , 3.11.8  & \xspace upsight-api.com & AndroidId \\
			\xspace com.yelp.android 9.5.1 &\xspace bugsnag.com  & AndroidId \\
			\xspace com.zeptolab.ctr.ads 2.8.0  &\xspace onesignal.com & AndroidId \\
			\xspace com.namcobandai\-games.pacmantournaments 6.3.0  &\xspace  namcowireless.com & AndroidId \\
			\xspace com.huuuge.casino.slots 2.3.185 &\xspace upsight-api.com & AndroidId \\
			\xspace com.cmplay.dancingline 1.1.1  &\xspace pushwoosh.com  & AndroidId\\
			\hline
		\end{tabular}
	}
	\caption{Applications with "jailbroken" field}
	\label{tab:jaibroken_apps}
\end{table}

{\bf Checking for Rooted Devices}. We noticed a suspicious flag  called ``jailbroken'' or ``device.jail\-broken'' leaked by several apps (\eg com.bitstrips.imoji, com.yelp.android, \zeptolab, etc). This flag was found in the URI content or in the body of a POST method  in the packets, and it was set to  1 if the device was rooted, or to 0 otherwise. In Table \ref{tab:jaibroken_apps}, we show the applications that contain this field in our dataset and the domain to which the ``jailbroken'' flag is being sent. We also show other types of leaks that the particular domain collects. From the table, we see that the flag is usually accompanied with a device identifier. Several apps send this flag to the same domain (\textvtt{upsight-api.com},  an ad network), which indicates that an ad library is probably leaking this information, rather than the app itself.

{\bf Behavioral Analysis of PII Leaks.}  An interesting direction for analyzing the \antShield dataset is via behavioral analysis. For instance, we can ask: (i) what can the communication between mobile apps and destination domains reveal about tracking and advertising? (ii) what type of information leaks to what domains and how to define similarity of apps or domains with respect to leaks? Fig. \ref{fig:graph} showcases one graph that visualizes similar destination domains with respect to leaks they received, as captured in the \antShield dataset.
We define  two domains to be similar if they are contacted by the same set of applications (see the box on the right inside Fig. \ref{fig:graph}).
For example, domains A and B are similar because they are contacted by two apps (app1, app2).
We depict the similarity of domains A and B as an edge on the graph of domains, at the bottom of the box.
This data can be readily extracted from our trace, together with the type of information that was transmitted from apps to domains.

The graph depicted on the left side of Fig. \ref{fig:graph} shows a projection of the underlying bipartite graph (middle step in the box) on domains (last step in the box); the graph is plotted  and analyzed using Gephi \cite{bastian2009gephi}. Nodes in this graph represent domains; the edges indicate similar nodes as per above definition; the width of the edge indicates the number of common applications; and the domain color corresponds to the type of leaked PII. The clusters of domains in the graph are the output of a community detection algorithm, which is a heuristic that tries to optimize modularity.\footnote{The main idea is that for specific node $i$, it tries to assign different communities of its neighbors like node $j$'s community as $i$'s community and compute the gain of modularity for whole network. The community which maximize the modularity will be the proper one. If the gain of modularity be negative or zero, $i$ keeps its community. This process is an iterative process which is done for all nodes. This algorithm is implemented in Gephi software \cite{bastian2009gephi}, and works with weighted graphs also.}

The graph in Figure \ref{fig:graph}  reveals  interesting patterns about PII leakage in the \antShield dataset.
First, advertising is the result of coordinated behavior.
For example, it is easy to identify ad exchanges: \textvtt{mopub.com} is in the center of all communication; and \textvtt{inner-active.mobi} and \textvtt{nexage.com} are also clearly shown as hubs.
All three large communities on the bottom and left of the graph correspond to ad networks.
Second, on the top left, there is a community of domains that belong mostly to Google and Facebook, and two domains (\textvtt{pof.com} and \textvtt{plentyoffish.com}) of a dating service.
The latter could be because the dating app also sends statistics (\eg for advertising purposes) to Google and Facebook, in addition to its own servers, as suggested by the type of PII being sent (gender and device ID, represented by the yellow color).
Third, not all domains belong to a community: some are well-behaved and are contacted only by their own app.
For instance the white-colored domain \textvtt{zillow.com} towards the bottom center of the graph is an isolate node and only receives information about the user's location, which makes sense since it provides a real-estate service.
Another example is the blue-colored domain \textvtt{hbonow.com}: it is only contacted by its own app and only receives the advertising ID to serve ads.
Another observation from Figure \ref{fig:graph} is that most domains in the same community receive the same type of PII (as indicated by the domain color).
This can be explained by the common ad libraries shared among different apps that fetch the same PII.

In general, similarity of apps and domains based on their network activity can be exploited to infer abusive behavior (\eg advertising, tracking, or malware) in mobile traffic, and this is one promising direction for future work.
\subsection{Classification Evaluation} \label{sec:prediction-result}

{\bf Classification Schemes under Comparison.} In this section, we use our datasets to compare the classification accuracy of the proposed \antShield approach (see Section \ref{sec:system-implementation}) to the previous state-of-the-art \recon approach (Section \ref{sec:related}). Since our proposed method combines several ideas, we also report results from the evaluation of individual ideas, to help assess which idea  brings the most benefit:
\begin{enumerate}
	\item Complete \recon approach as per Section \ref{sec:related}: classify all  (\preDefLeak and \unknownLeak) exposures, using binary classifiers first to detect a leak, then heuristics to determine the type of leak. %
	\item \recon classifying \unknownLeak exposures only.
	\item \stringMatch on \preDefLeak exposures, \recon trained on \unknownLeak; testing done on all exposures.
	\item \multiLabel  classification trained and tested on \preDefLeak and \unknownLeak exposures.
	\item \multiLabel  classification trained and tested only on \unknownLeak.
	\item Complete \antshield as per Section \ref{sec:system-implementation}: \stringMatch for \preDefLeak and \multiLabel classification  for \unknownLeak leaks;  \multiLabel trained on \unknownLeak only, testing done on all exposures.
	
\end{enumerate}

\begin{table*}[t!]
	\centering
	{\footnotesize
		\begin{tabular}{cc|c|p{2cm}|p{2cm}|p{2cm}|p{2cm}}
			\cline{3-6}
			& & \multicolumn{4}{ c| }{Method} \\ \cline{3-6}
			&                                   & \recon on All PII & \recon on \unknownLeak & \multiLabel on All PII &  \multiLabel on \unknownLeak \\ \cline{1-6}
			\multicolumn{1}{ |c  }{\multirow{3}{*}{\parbox{2.5cm}{Per-Domain Average}} } &
			\multicolumn{1}{ |c| }{F-measure}   & { \bf 98.0\% $\pm$ 6.80 } & 97.2\% $\pm$ 14.2      & 97.1\% $\pm$ 9.32      & { \bf 97.2\% $\pm$ 11.3 }  &\\ \cline{2-6}
			\multicolumn{1}{ |c  }{}            &
			\multicolumn{1}{ |c| }{specificity} & { \bf 97.5\% $\pm$ 8.10 } & 98.5\% $\pm$ 4.35      & 98.4\% $\pm$ 5.75      & { \bf 98.9\% $\pm$ 5.72 }  &\\ \cline{2-6}
			\multicolumn{1}{ |c  }{}            &
			\multicolumn{1}{ |c| }{recall}      & { \bf 98.5\% $\pm$ 6.78 } & 97.9\% $\pm$ 14.1      & 97.1\% $\pm$ 9.38      & { \bf 97.3\% $\pm$ 9.46 }  &\\ \cline{1-6}
			
			\multicolumn{1}{ |c  }{\multirow{3}{*}{\parbox{2.5cm}{Per-App Average}} } &
			\multicolumn{1}{ |c| }{F-measure}   & { \bf 97.0\% $\pm$ 7.99 } & 96.4\% $\pm$ 14.9      & 96.2\% $\pm$ 7.25      & { \bf 96.4\% $\pm$ 12.0 }  & \\ \cline{2-6}
			\multicolumn{1}{ |c  }{}            &
			\multicolumn{1}{ |c| }{specificity} & { \bf 98.1\% $\pm$ 4.24 } & 96.8\% $\pm$ 11.3      & 96.4\% $\pm$ 7.30      & { \bf 98.3\% $\pm$ 8.87 }  & \\ \cline{2-6}
			\multicolumn{1}{ |c  }{}            &
			\multicolumn{1}{ |c| }{recall}      & { \bf 96.4\% $\pm$ 8.95 } & 97.6\% $\pm$ 14.5      & 97.4\% $\pm$ 6.48      & { \bf 95.9\% $\pm$ 12.1 }  & \\ \cline{1-6}
			
			\multicolumn{1}{ |c  }{\multirow{3}{*}{\parbox{2.5cm}{General}} } &
			\multicolumn{1}{ |c| }{F-measure}   & { \bf 97.5\% }           & 94.9\%                 & 95.5\%                 & { \bf 99.5\% }  &\\ \cline{2-6}
			\multicolumn{1}{ |c  }{}            &
			\multicolumn{1}{ |c| }{specificity} & { \bf 98.9\% }           & 99.8\%                 & 95.6\%                 & { \bf 91.8\% }  &\\ \cline{2-6}
			\multicolumn{1}{ |c  }{}            &
			\multicolumn{1}{ |c| }{recall}      & { \bf 95.8\% }           & 91.9\%                 & 98.4\%                 & { \bf 99.8\% }  & \\ \cline{1-6}
			
		\end{tabular}
	}
	\caption{Binary Classification Results (Sec. \ref{sec:prediction-binary})} \label{tbl:eval-binary}
\end{table*}

\begin{table*}[t!]
\centering

	{\footnotesize
	\begin{tabular}{cc|p{2cm}|p{2cm}|p{2cm}|p{2cm}|p{2cm}|p{2cm}|p{2cm}}
		\cline{3-8}
		& & \multicolumn{6}{ c| }{Method} \\ \cline{3-8}
		& & (1) \recon on All PII   & (2) \recon on \unknownLeak & (3) \stringMatch \& \recon on \unknownLeak & (4) \multiLabel on All PII & 
		(5) \multiLabel on \unknownLeak & (6) \stringMatch \& \multiLabel \\ \cline{1-8}
		\multicolumn{1}{ |c  }{\multirow{3}{*}{\parbox{1.5cm}{Per-Domain Avg}} } &
		\multicolumn{1}{ |c| }{accuracy}  & { \bf 72.7\% $\pm$ 39.7 } & 69.5\% $\pm$ 45.5     & 95.9\% $\pm$ 18.4                      & 99.2\% $\pm$ 1.90      & 
		99.3\% $\pm$ 2.88      & { \bf 98.5\% $\pm$ 11.0 }  &    \\ \cline{2-8}
		\multicolumn{1}{ |c  }{}          &
		\multicolumn{1}{ |c| }{precision} & { \bf 74.8\% $\pm$ 39.3 } & 69.5\% $\pm$ 45.5     & 96.2\% $\pm$ 18.1                      & 99.3\% $\pm$ 1.95      & 
		99.3\% $\pm$ 3.21      & { \bf 98.5\% $\pm$ 11.0 }  &    \\ \cline{2-8}
		\multicolumn{1}{ |c  }{}                        &
		\multicolumn{1}{ |c| }{recall}    & { \bf 73.5\% $\pm$ 39.6 } & 69.5\% $\pm$ 45.5     & 95.9\% $\pm$ 18.4                      & 99.3\% $\pm$ 1.79      & 
		99.5\% $\pm$ 2.11      & { \bf 98.9\% $\pm$ 10.4 } &    \\ \cline{1-8}
		
		\multicolumn{1}{ |c  }{\multirow{3}{*}{\parbox{1.5cm}{Per-App Avg}} } &
		\multicolumn{1}{ |c| }{accuracy}  & { \bf 73.2\% $\pm$ 31.1 } & 69.0\% $\pm$ 42.7      & 97.6\% $\pm$ 13.1                      & 98.8\% $\pm$ 2.24     & 
		98.9\% $\pm$ 3.23      & { \bf 99.4\% $\pm$ 4.58 }   &    \\ \cline{2-8}
		\multicolumn{1}{ |c  }{}          &
		\multicolumn{1}{ |c| }{precision} & { \bf 76.7\% $\pm$ 30.4 } & 69.0\% $\pm$ 42.7      & 98.0\% $\pm$ 12.8                      & 98.9\% $\pm$ 2.20     & 
		99.0\% $\pm$ 3.29     & { \bf 99.4\% $\pm$ 4.58 }  &    \\ \cline{2-8}
		\multicolumn{1}{ |c  }{}                        &
		\multicolumn{1}{ |c| }{recall}    & { \bf 73.5\% $\pm$ 31.0 } & 69.1\% $\pm$ 42.8      & 97.6\% $\pm$ 13.1                      & 98.9\% $\pm$ 2.18     & 
		99.1\% $\pm$ 2.40     & { \bf 100\% $\pm$ 0.06 }    &    \\ \cline{1-8}
		
		\multicolumn{1}{ |c  }{\multirow{3}{*}{\parbox{1.5cm}{General}} } &
		\multicolumn{1}{ |c| }{accuracy}  & { \bf 49.9\% }           & 50.2\%                 & 97.1\%                                 & 77.4\%                & 
		81.8\%                      & { \bf 99.3\% }         &    \\ \cline{2-8}
		\multicolumn{1}{ |c  }{}          &
		\multicolumn{1}{ |c| }{precision} & { \bf 58.2\% }          & 50.3\%                 & 97.6\%                                 & 79.6\%                & 
		84.7\%                     & { \bf 99.5\%  }        &    \\ \cline{2-8}
		\multicolumn{1}{ |c  }{}                        &
		\multicolumn{1}{ |c| }{recall}    & { \bf 53.3\% }           & 50.3\%                 & 97.1\%                                 & 75.9\%                & 
		79.4\%                     & { \bf 99.7\%  }        &    \\ \cline{1-8}
		
	\end{tabular}
}
\caption{Leak Classification Results (Sec. \ref{sec:prediction-leak}).} \label{tbl:eval-leak}
\end{table*}

\begin{table*}[t!]
\centering

{\footnotesize
	\begin{tabular}{cc|p{2cm}|p{2cm}|p{2cm}|p{2cm}|p{2cm}|p{2cm}|p{2cm}}
		\cline{3-8}
		& & \multicolumn{6}{ c| }{Method} \\ \cline{3-8}
		                                  & & (1) \recon on All PII & (2) \recon on \unknownLeak & (3) \stringMatch \& \recon on \unknownLeak & (4) \multiLabel on All PII & 
		                                      (5) \multiLabel on \unknownLeak & (6) Complete \antShield \\ \cline{1-8}
		\multicolumn{1}{ |c  }{\multirow{3}{*}{\parbox{1.5cm}{Per-Domain Avg}} } &
		\multicolumn{1}{ |c| }{accuracy}  & { \bf 89.1\% $\pm$ 22.1}   & 91.8\% $\pm$ 17.7      & 98.5\% $\pm$ 7.81                      & 99.2\% $\pm$ 2.02      & 
		                                  99.3\% $\pm$ 2.54               &{ \bf 99.5\% $\pm$ 3.99}        &    \\ \cline{2-8}
		\multicolumn{1}{ |c  }{}          &
		\multicolumn{1}{ |c| }{precision} & { \bf 90.0\% $\pm$ 21.5}   & 91.8\% $\pm$ 17.7      & 98.7\% $\pm$ 7.55                      & 99.2\% $\pm$ 2.10      & 
		                                  99.3\% $\pm$ 2.82               &{ \bf 99.5\% $\pm$ 3.99}        &    \\ \cline{2-8}
		\multicolumn{1}{ |c  }{}                        &
		\multicolumn{1}{ |c| }{recall}    & { \bf 89.2\% $\pm$ 22.0}   & 91.8\% $\pm$ 17.7      & 98.5\% $\pm$ 7.80                      & 99.2\% $\pm$ 1.83      & 
		                                  99.5\% $\pm$ 1.87     &{ \bf 99.8\% $\pm$ 1.60}                  &    \\ \cline{1-8}
		
		\multicolumn{1}{ |c  }{\multirow{3}{*}{\parbox{1.5cm}{Per-App Avg}} } &
		\multicolumn{1}{ |c| }{accuracy} & { \bf 91.3\% $\pm$ 15.2}   & 95.0\% $\pm$ 12.6      & 99.5\% $\pm$ 3.09                      & 98.7\% $\pm$ 2.31       & 
		                                 98.9\% $\pm$ 2.83               &{ \bf 99.1\% $\pm$ 7.35}        &    \\ \cline{2-8}
		\multicolumn{1}{ |c  }{}          &
		\multicolumn{1}{ |c| }{precision} &{ \bf 92.7\% $\pm$ 14.1}  & 95.0\% $\pm$ 12.6      & 99.5\% $\pm$ 2.96                      & 98.7\% $\pm$ 2.24       & 
		                                 99.0\% $\pm$ 2.89               &{ \bf 99.1\% $\pm$ 7.35}        &    \\ \cline{2-8}
		\multicolumn{1}{ |c  }{}                        &
		\multicolumn{1}{ |c| }{recall}    &{ \bf 91.4\% $\pm$ 15.2}  & 95.0\% $\pm$ 12.6      & 99.5\% $\pm$ 3.06                      & 98.7\% $\pm$ 2.26       & 
		                                 99.1\% $\pm$ 2.11               &{ \bf 99.4\% $\pm$ 6.91}        &    \\ \cline{1-8}
		                                 
		\multicolumn{1}{ |c  }{\multirow{3}{*}{\parbox{1.5cm}{General}} } &
		\multicolumn{1}{ |c| }{accuracy} &{ \bf 89.8\%}              & 99.1\%                 & 99.3\%                                 & 78.5\%                  & 
		76.5\%                      &{ \bf 99.8\%}           &    \\ \cline{2-8}
		\multicolumn{1}{ |c  }{}          &
		\multicolumn{1}{ |c| }{precision} &{ \bf 91.3\%}             & 99.1\%                 & 99.4\%                                 & 80.6\%                  & 
		79.1\%                     &{ \bf 99.8\%}           &    \\ \cline{2-8}
		\multicolumn{1}{ |c  }{}                        &
		\multicolumn{1}{ |c| }{recall}    &{ \bf 90.4\%}             & 99.1\%                 & 99.4\%                                 & 77.0\%                  & 
		74.4\%                     &{ \bf 99.9\%}           &    \\ \cline{1-8}

	\end{tabular}
}
\caption{Combined Classification Results (Sec. \ref{sec:prediction-combined}).}
 \label{tbl:eval-combined}
\end{table*}

{\bf Per-app vs. Per-domain classifiers.} In Section \ref{sec:system-implementation}, we discussed the system advantages of using per-app instead of per-domain classifiers. In this section, we show that their classification performance is similar (which is also justified by the insights at the end of Section \ref{sec:findings}). For each method, we compare how well the per-domain, per-app, and general classifiers perform.  We train specialized classifiers  for those domains and apps that contain at least one positive sample (packet with an exposure), and one negative sample (packet with no exposure).  In that sense, we find that per-app classifiers are able to cover more data than the per-domain classifiers. In particular, we obtain the following numbers for packets covered by a classifier:

\begin{itemize}
	\item All PII, per-app classifiers: 211 (93.3\% of traffic, 99.5\% of packets with PII) 
	\item All PII, per-domain classifiers: 182 (63.6\% of traffic, 95.0\% of packets with PII)
	\item \unknownLeak PII, per-app classifiers: 47 (54.4\% of traffic, 99.5\% of packets with \unknownLeak PII) 
	\item \unknownLeak PII, per-domain classifiers: 49 (24.5\% of traffic, 87.4\% of packets with \unknownLeak PII)
\end{itemize}

This is expected since apps generally exhibit more diverse behavior by connecting to various domains, some of which collect PII and some of which do not. Thus, we are more likely to find apps that have sent at least one packet containing PII and one packet without PII, as opposed to domains that receive packets with and without PII.

\begin{figure}[t!]
	\centering
	\includegraphics[width=0.5\textwidth]{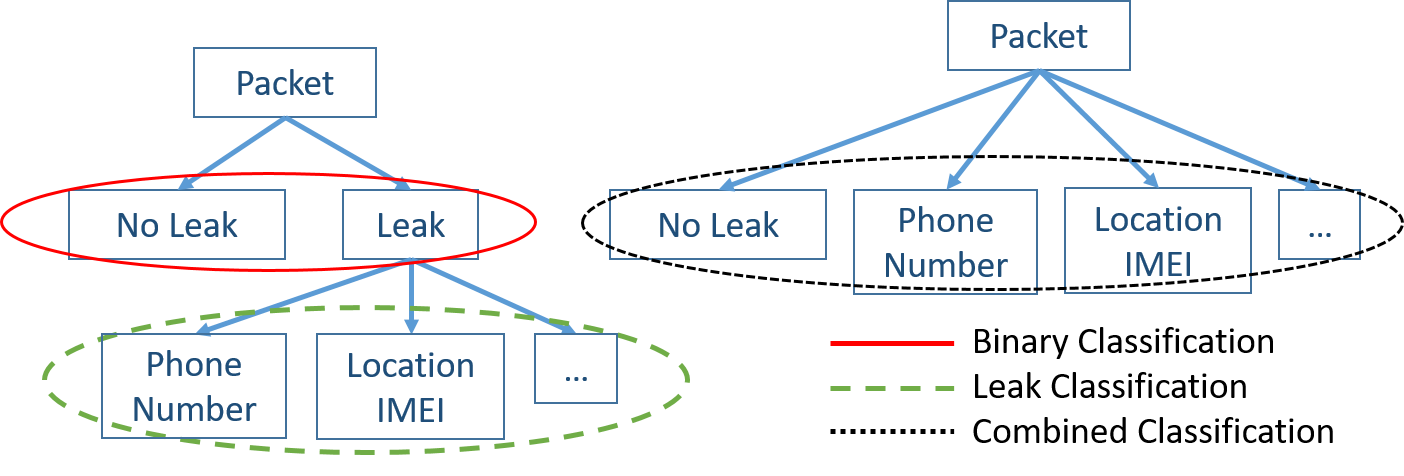}
	\caption{{\footnotesize Evaluation approaches: (1) Binary Classification: we assess how well we identify whether or not a packet contains a PII (Sec. \ref{sec:prediction-binary}); (2) Leak Classification: we assess how well we infer the  PII type from packets that already contain a PII, ignoring packets without PII (Sec. \ref{sec:prediction-leak}); (3) Combined Classification - assess how well we identify the PII type {\em and} the No Leak label, considering all packets (Sec. \ref{sec:prediction-combined}).}
	}\label{fig:evalTrees}
	\vspace{-10pt}
\end{figure}

{\bf Evaluation Approaches and Metrics.} After classifying a packet, either a leak is detected with a particular PII type, or No Leak is detected. Depending on how one summarizes these numbers over all packets classified, we may have different assessments. In particular, whether or not we consider packets that do not contain PII, affects the numbers, since this is the majority of the packets. We considered three evaluation schemes, summarized in Fig. \ref{fig:evalTrees}:
\begin{enumerate}
	\item {\em Binary Classification}: this approach evaluates how well the applicable algorithms classify a packet as containing an exposure or not (Sec. \ref{sec:prediction-binary}).
	\item {\em Leak Classification}: this approach evaluates how well each algorithm distinguishes PII types in packets that contain an exposure (Sec. \ref{sec:prediction-leak}), \ie packets without a PII are not taken into account.  	%
	\item {\em Combined Classification}: this approach evaluates how well each algorithms distinguishes among PII types and ``no leak'' (Sec. \ref{sec:prediction-combined}), \ie packets without a PII are taken into account.
\end{enumerate}
For each approach, we perform 5-fold cross-validation on the given model (unless otherwise specified), and calculate the average and the standard deviation across the trained specialized  classifiers. (Since \recon's second (non-binary) step and \stringMatch are both heuristic, we did not perform cross-validation on these methods when evaluating leak and combined classification, but simply ran the algorithms on the entire applicable dataset (columns 1-3 and 6) in the Tables.)

Because a packet can leak more than one PII type, for the latter two approaches, we use evaluation metrics specific to multi-labeling problems \cite{godbole2004discriminative}. We report : (i) {\em accuracy}: the number of correct labels, divided by the number of predicted and true labels, (ii) {\em precision}: the number of correct labels, divided by the number  of predicted labels, (iii) {\em recall}: the number of correct labels, divided by the number  of true labels.
\subsubsection{Binary Classification} \label{sec:prediction-binary}
We report the binary classification results in Table \ref{tbl:eval-binary} for the two machine learning algorithms under consideration: \recon's DT, and our \multiLabel BR. We report the standard metrics for binary classification: F-measure, specificity, and recall. The first column is consistent with \recon's own reports in \cite{recon15} - the model achieves high accuracy and low false positives/negatives. The second column shows that there is little benefit in focusing on \unknownLeak exposures only. This makes sense, since in this binary step, we only want to see whether or not a packet contains a leak, and not to extract what type of leak it is (see Fig. \ref{fig:evalTrees}). The third and fourth columns also show little benefit from our \multiLabel approach since within the BR, we still use a similar decision tree to classify exposure vs non-exposure. We also note that the standard deviation is higher when focusing on \unknownLeak exposures only (columns 2 and 4). This is expected since there is now less data to work with and some domains send \unknownLeak PII only once in a while. Furthermore, in the case of binary classification, the general classifiers perform close to the specialized ones. However, we are interested in improving the accuracy on the type of PII classification, and as we show in the next two subsections, our approaches and the specialized classifiers bring benefit there.

\subsubsection{Leak Classification} \label{sec:prediction-leak}

The main results are summarized in Table \ref{tbl:eval-leak}. First, standard deviation is high because certain domains are easy to learn and get near 100\%, while a small set of domains are difficult (some even have 0\% accuracy).  \recon's heuristic scores low when attempting to extract the PII type (column 1); see Sec. \ref{sec:related} and \cite{recon15} for a description of the heuristic. Second, when we reduce the set of PII types to look for (column 2), the heuristic performs slightly worse, probably due to not having enough samples of \unknownLeak exposures. Third, as expected, \stringMatch can find \preDefLeak exposures  with 100\% accuracy, thus the overall accuracy improves by \customtilde20\% (column 3 vs column 1), and standard deviation decreases. Fourth, the \multiLabel approach shows significant improvement when compared to \recon's heuristic (column 4 vs column 1, and column 5 vs column 2); this is expected, since we do not need to estimate probabilities or calculate out thresholds. Fifth, the complete \antShield achieves near perfect performance, and decreases the standard deviation (column 6 vs columns 1-3). %
 Finally, in all cases: (i) the specialized classifiers outperform the general ones, and (ii) the per-app classifiers achieve higher accuracy and lower standard deviation in our final method (column 6).

\subsubsection{Combined Classification} \label{sec:prediction-combined}
The results for combined classification are shown in Table \ref{tbl:eval-combined} and the difference between the performance of different classification methods is less pronounced than before. This is because the majority of packets do not contain a leak, the binary classifiers work well (see Sec. \ref{sec:prediction-binary}) and classify the "no exposure" packets correctly, making the results look deceivingly good. This is why we also report the Leak Classification performance (Sec. \ref{sec:prediction-leak}), as it provides deeper insight into the classifiers' performance. We note that in this case, the \multiLabel general classifiers appear to do worse than the corresponding ones in \recon, because the results reported in columns 4 and 5 are based on cross-validation, so the general classifiers do not see all the training data and do worse on some folds.

\subsection{Real-Time Performance on the Device} \label{sec:eval-time}
In order to run  privacy leakage detection in real-time on the device, performance is key. Thus, we evaluate the two feature extraction approaches: (1) \recon's Java string parsing, and (2) \antLib's Aho-Corasick search for features and \preDefLeak PII. We also compare: (1) \recon's binary classification, and (2) \antShield's \multiLabel classification. We find that our classifiers have negligible impact on battery and can run in real-time. This is mainly thanks to the use of (i) Aho-Corasik for searching for multiple strings, and (ii) the lean extraction of words to feed into the classifiers. To the best of our knowledge, this is the first time that PII classification is achieved in real-time on a mobile device.

{\bf Setup.}
The tests were ran on a Nexus 6P with Android 7.1.1 and an 8-core QUALCOMM Snapdragon 810 processor with a clock speed of 2 GHz and battery capacity of 3450 mAh. %
We fed 10 HTTP packets of varying sizes (between 300-2000B) to each function under evaluation and timed how long it took using \textvtt{System.nanoTime()}. We repeated each test case 100 times and calculated the average run-time and standard deviation. Each function was tested in isolation, running on the main thread, so as to minimize timing the overhead of possible thread switching.

{\bf Results.}
The results for the feature extraction approaches are as follows: (1) \recon's Java string parsing: 36 ms $\pm$ 17 ms; (2) Aho-Corasick search: 0.107 ms $\pm$ 0.149 ms. Clearly, \antLib's efficient Aho-Corasick implementation brings orders of magnitude of benefit. %

The results for classification techniques are: (1) \recon's binary classification: 0.041 ms $\pm$ 0.029 ms; (2) \multiLabel classification: 0.751 ms $\pm$ 1.35 ms. As expected, the \multiLabel classification takes a little longer, but it is still reasonable  and will not significantly impact user experience.

{\bf Training Time.} Training \multiLabel classifiers generally takes twice as long as \recon's binary classifiers. However, in both cases, a specialized classifier is trained within tens of {\em milliseconds} even when done on a standard Windows 10 laptop. General classifiers can take up to tens of minutes. %
 When considering only \unknownLeak leaks, the number of labels is reduced, and both the binary and the \multiLabel general classifiers take {\em under 10 min} to train. However, since training is performed infrequently, and can be done at a remote server (the classifiers can be fetched later by user devices), we consider the training times a non-issue.

\section{Conclusion \& Future Directions} \label{sec:conclusion}

We presented  \antShield\xspace- a system that performs, for the first time, on-device detection of \preDefLeak PII and classification of \unknownLeak PII, accurately (with higher accuracy and lower variance than state-of-the-art) and with low overhead (in real-time on the device). In the process, we collected and analyzed a new dataset, which reveals  interesting PII leaks and patterns, some of which were previously unknown. Preliminary graph analysis revealed interesting patterns of apps and domains colluding to leak private information; behavioral analysis of PII leaks is one promising direction for future work. 
 We  will make the \antShield work available to the research community, including the \antShield plugin for the \antLib on the device, the training module, and the \antShield dataset. 

 This work focused on a single device and it is the first necessary step towards enabling distributed learning of PII leakage, where multiple devices running \antShield  collaborate with each other and/or a central entity to share training data and/or classifiers. This is an important direction for future work.
If the users do not completely trust the central entity, distributed machine learning frameworks for enabling collaborative learning, while preserving user privacy, are currently an active research area.  We will consider Federated Learning  \cite{fedlearn,Bonawitz} (which enables mobile phones to collaboratively learn a shared prediction model while keeping all the training data on device, and is supported by Google); the Teacher-Student model  \cite{papernot2017semi} (ensemble-based machine learning on private datasets); and hybrid approaches such as Blender \cite{blender} (a hybrid differential privacy model where users have different privacy requirements).

\bibliographystyle{unsrt}

\end{document}